\documentclass[runningheads]{llncs}
\usepackage{graphicx}
\usepackage{algorithm}
\usepackage{algorithmic}
\usepackage{textcomp}
\usepackage{amsmath}
\usepackage{mathrsfs}
\usepackage{multirow} 
\usepackage{diagbox} 
\usepackage{xcolor}
\usepackage{bm}
\usepackage{makecell}
\usepackage{subfig}
\usepackage[misc]{ifsym}
\begin{document}
\title{RecGPT: Generative Personalized Prompts for Sequential Recommendation via ChatGPT Training Paradigm}

\author{Yabin Zhang\inst{1}, Wenhui Yu\textsuperscript{2,(\Letter)}, Erhan Zhang\inst{3}, Xu Chen\inst{1}, Lantao Hu\inst{2},\\ Peng Jiang\inst{2}, Kun Gai\inst{4}}
\institute{Gaoling School of Artificial Intelligence, Renmin University of China, China\\
	\email{\{yabin.zhang,xu.chen\}@ruc.edu.cn}
\and Kuaishou Technology, Beijing, China \\
\email{yuwenhui07@kuaishou.com,hulantao@gmail.com,jp2006@139.com}
\and Peking University, Beijing, China,\\
\email{zhangeh-ss@stu.pku.edu.cn};
\and Unaffliated, Beijing, China,\\
\email{\{gai.kun\}@qq.com}
}

\maketitle              
\begin{abstract}
ChatGPT has achieved remarkable success in natural language understanding. Considering that recommendation is indeed a conversation between users and the system with items as words, which has similar underlying pattern with ChatGPT, we design a new chat framework in item index level for the recommendation task. Our novelty mainly contains three parts: model, training and inference. For the model part, we adopt Generative Pre-training Transformer (GPT) as the sequential recommendation model and design a user modular to capture personalized information. For the training part, we adopt the two-stage paradigm of ChatGPT, including pre-training and fine-tuning. In the pre-training stage, we train GPT model by auto-regression. In the fine-tuning stage, we train the model with prompts, which include both the newly-generated results from the model and the user's feedback. For the inference part, we predict several user interests as user representations in an autoregressive manner. For each interest vector, we recall several items with the highest similarity and merge the items recalled by all interest vectors into the final result. We conduct experiments with both offline public datasets and online A/B test to demonstrate the effectiveness of our proposed method.
\end{abstract}

\section{Introduction}
Recently, ChatGPT has gained significant attention due to the impressive capabilities and unique training paradigm \cite{radford2018improving}. Moreover, it has found extensive practical application in diverse domains such as law \cite{choi2023chatgpt}, ethics \cite{shen2023chatgpt}, education \cite{khalil2023will} and reasoning \cite{bang2023multitask}. This has motivated researchers to explore its potential application in recommendation systems. 

In the past year, many recommendation models based on ChatGPT have been proposed. For example, \cite{gao2023chat} proposed Chat-REC by leveraging ChatGPT to provide transparent and understandable reasoning for recommendations, thereby fostering trust and enhancing user engagement. \cite{li2023gpt4rec} proposed a generative framework called GPT4Rec, that generates personalized recommendations and interpretable user interest representations simultaneously. \cite{wang2023zero} proposed a prompting strategy that guides GPT-3 in generating next-item recommendations. These methods suffer from the following limitations: (1) In these methods, users and items are represented in the semantic space, which shows limitation in representing the complex preference relationship. (2) These large language models are too heavy for online recommendation services. Therefore, exploring flexible and capable methods of integrating ChatGPT into recommendation systems has become a promising direction. 

\begin{figure}[t]
	\centerline{{\includegraphics[width=3.5in,height=2.3in]{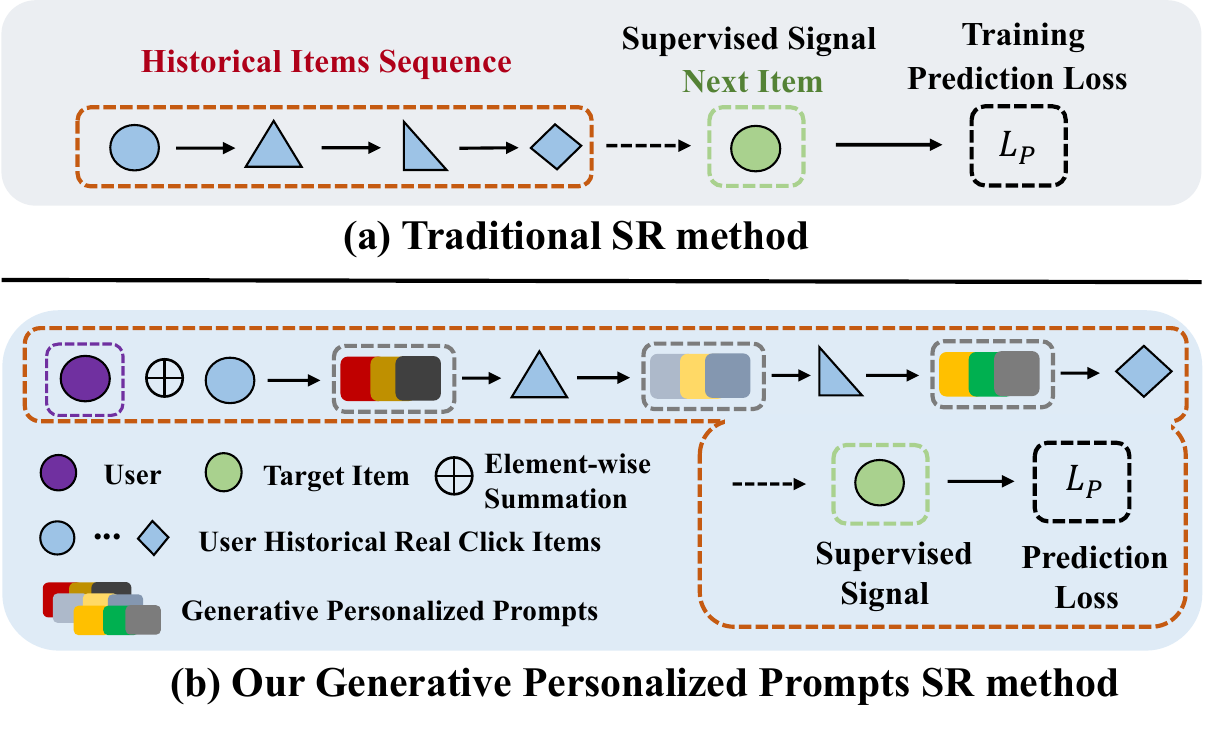}}} 
	\vspace*{-0.2cm}
	\caption{Illustration of training mechanisms of different methods for sequential recommendation. (a) Traditional sequential recommendation methods. (b) Generative personalized prompts for sequential recommendation.}
	\vspace*{-0.5cm}
	\label{figure1}
\end{figure}
In this paper, we abandon the form of natural language, we use the model structure and training paradigm of ChatGPT for item sequence prediction. When chatting with a chatbot such as ChatGPT, a user asks a question and the model returns the answer. He/she then gives reaction to the answer and the model returns the answer again. All questions and answers are sequences of words. Similarly, when browsing items on the web, a user request the model with her historical action list and the model returns an item list. She then gives the feedback (clicked item list) to the model and the model returns a new recommendation list based on the feedback. As we can see, recommendation is to communicate to the user with items as tokens. Considering the underlying similarity, we transfer the technique of ChatGPT into sequential recommendation. sequential recommendation (SR) aims to predict users' following preferences based on historical behavior sequences \cite{kang2018self,wu2020sse,hyun2022beyond,zhou2022filter} and is a key component of most online recommendation scenarios. Although many methods have been proposed for SR, most of them only consider to model via users' historical behavior sequence, as depicted in Fig. 1 (a). However, these behavior sequences are highly sparse compared to the entire interaction space. Additionally, there is a gap between this approach and real-world online recommendation systems. It is common practice to recommend multiple items to users, who subsequently provide feedback for each item. For example, a recommender system might suggest six items to a user, but only one may be clicked. In current sequential models, the focus lies on the users' clicked items, assigned a confidence level of 100\%. In contrast, the confidence levels of remaining five unclicked items, which can range from 0\% to 100\%, are not taken into account. Unlike modeling only using the user's behavioral sequence, we propose incorporating unclicked items as personalized prompts in the training of models, as shown in Fig. 1(b). This method introduces the migration of user preferences over time.

To enhance personalization of prompts in ChatGPT-based recommendation models and mitigate the limitation of modeling a single behavior sequence in traditional sequential scenarios, in this paper, we proposes a new framework called \textbf{RecGPT} (Generative Personalized Prompts for Sequential \textbf{Rec}ommendation via Chat\textbf{GPT} training paradigm). RecGPT utilizes generative pre-training model to generate various prompts besides the original user behavior sequence items. In accordance with the ChatGPT training paradigm, we employ a training procedure consisting of two stages: Pre-training and fine-tuning. In the Pre-training stage, we pre-train a multi-layer Transformer decoder network model with auto-regressive generative capabilities. To enhance the model's personalized generation abilities, we integrate user IDs modular.  In the fine-tuning stage,  we have two parts including Prompt-tuning, and Inference-validation. Specifically, we first fine-tune the pre-trained model to generate personalized prompts specifically tailored for the recommendation task.  We introduce segment id in the model, which is used to distinguish the generated prompts from the original behavior sequence. Then, in the inference-validation stage, we substitute the conventional inner product recall method with an auto-regressive recall approach to assess the performance of recommendations. The auto-regressive recall approach is founded on the sequential output of multiple user vectors, effectively capturing the evolution of user preferences over time. Additionally, we have also incorporated the above method in four offline dataset and an online the recalling phase of the \textit{Kuaishou}\footnote{https://www.kuaishou.com/en} App's to demonstrate its effectiveness.

To summarize, this work makes the following main contributions:

$\bullet$ We propose to model user behavior sequences with a personalized prompt via the ChatGPT training paradigm, called RecGPT. 

$\bullet$ We implemented the above idea through three parts: first, we pre-trained a personalized auto-regressive generative model by introducing user IDs modular; then, we fine-tuned the pre-trained model by introducing segment IDs to generate personalized prompts specifically for the recommendation task; finally, we verified the evolution of user preferences over time using a two-step auto-regressive recall method during the inference stage.

$\bullet$ To the best of our knowledge, we are the first to employ auto-regressive recall in an online A/B test and demonstrate its viability. This provides a new perspective from ChatGPT on online recommendation, which helps motivate the following research in both the academic community and industry.

$\bullet$ We conduct extensive offline and an online A/B test experiments to demonstrate the effectiveness of RecGPT.

\section{Related Work}
\subsection{Sequential Recommendation} 
Sequential Recommendation aims to predict the next-item that best fits user preferences based on the user's historical interactions.  Pioneering works \cite{he2016fusing,rendle2010factorization} for sequential recommendation mainly model an item-to-item transaction pattern according to Markov Chains assumption. \cite{rendle2010factorizing}  is based on personalized transition graphs over underlying Markov chains, which model subsumes both a common Markov chain and matrix factorization. Recurrent Neural Networks (RNNs) further improved SR performance, with models like GRU4Rec \cite{hidasi2015session} and LSTM-based methods \cite{wu2017recurrent} capturing long-term and short-term item transition correlations. Owing to success of self-attention \cite{devlin2018bert,vaswani2017attention} models in NLP tasks, researches also designed quite a lot of models based on Transformer-based, such as SASRec \cite{kang2018self} brings self-attention into sequential, BERT4Rec \cite{sun2019bert4rec} employs BERT model to learn bidirectional item dependencies for sequential recommendation.  However, existing methods lack modeling a multi-feedback behavior sequence.

\subsection{ChatGPT for Recommendation}
ChatGPT is a recent chatbot service released by OpenAI and is receiving increasing attention over the past year. Recently, many ChatGPT-based and LLM-based \cite{wang2024survey} recommendation models have been proposed for various recommendation tasks, such as sequential recommendation \cite{wu2022personalized,li2023gpt4rec,wang2023zero} and explainable recommendation \cite{gao2023chat}. For instance, Chat-REC \cite{gao2023chat} leverages ChatGPT to offer transparent and understandable reasoning for its suggestions, thereby fostering trust and enhancing user engagement. GPT4Rec \cite{li2023gpt4rec} proposes a generative framework that generates personalized recommendations and interpretable user interest representations simultaneously. In \cite{wang2023zero}, a prompting strategy is proposed to guide GPT-3 in making next-item recommendations. However, these approaches are only implemented through the official OpenAI API and directly applying these models to ID-based online recommendation systems is not feasible. Unlike these methods, we introduce the ChatGPT training paradigm to the sequential recommendation task and demonstrate its feasibility.

\section{Preliminary}
Assume we have a set of users $u \in \mathcal U$ and a set of items $v \in \mathcal V$. For each user, $s_{u} = \{v_{u,1},\cdots,v_{u,j},\cdots,v_{u,|s_{u}|} \}$ denotes the sequence of user historical behaviors in chronological order, where  $v_{u,j}$ denotes the $j^{th}$ item interacted by the user and $|s_{u}|$ is the total number of items. Given an observed sequence $s_{u}$, the typical task of SR is to predict the next items $v_{u,|s_{u}|+1}$ that the user $u$ is most likely to be interacted with, which is formulated as follows:
\begin{equation}
	\begin{split}
		\underset {v_{i} \in \mathcal{V}} { \operatorname {arg\,max} } \, P (v_{|s_{u}|+1} = v_{i} | s_{u}), \\
	\end{split}
\end{equation}
which is interpreted as calculating the probability of all candidate items and selecting the highest one for the user.
\begin{figure*}[t]
	\centerline{{\includegraphics[width=4.7in,height=2.9in]{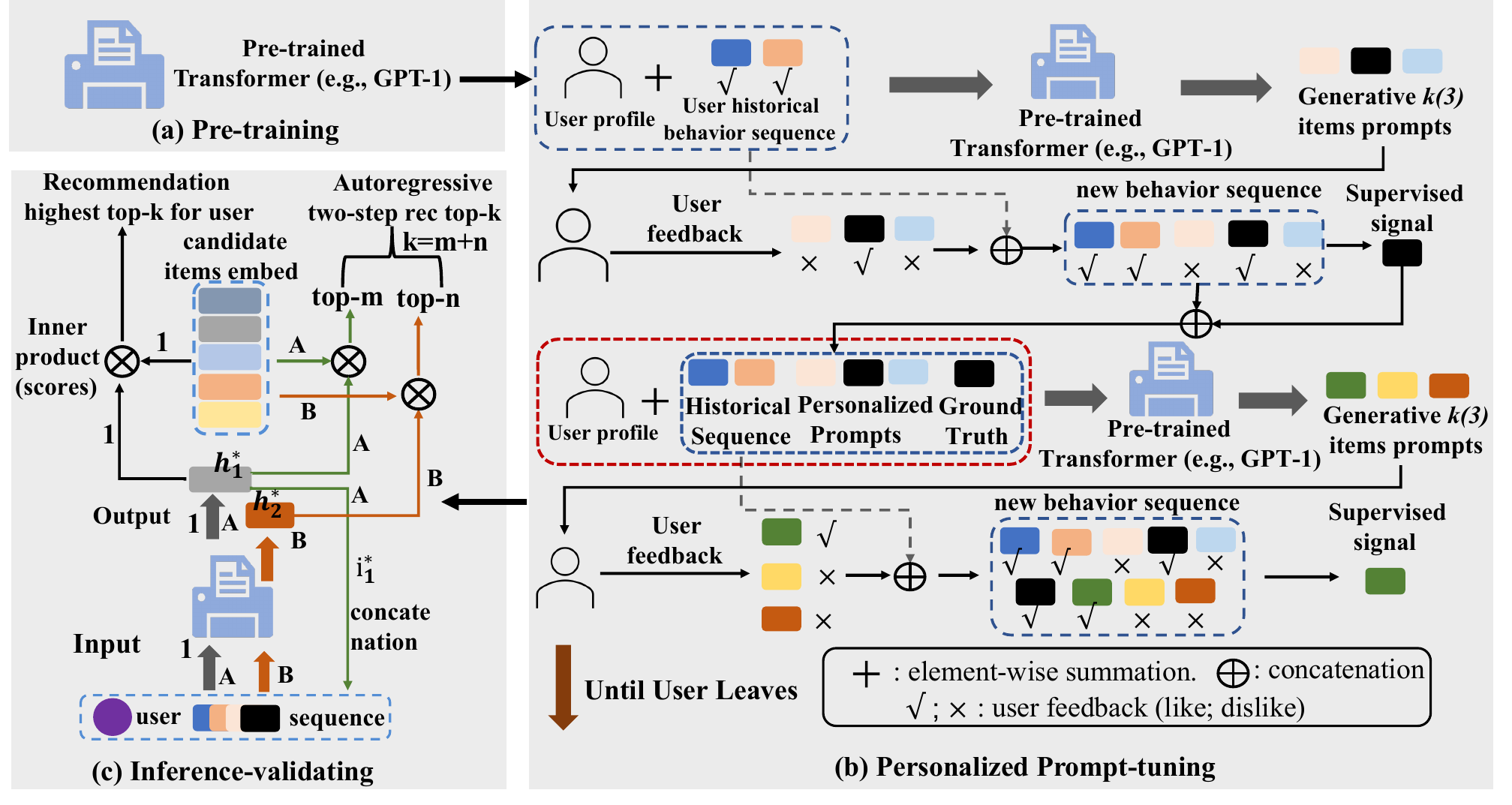}}} 
	\caption{Overview of our proposed RecGPT framework. (a) Pre-training stage. (b) Conversations between users and GPT by personalized prompts. (c) To validate auto-regressive generative recommendation ability. The path 1 is traditional inference method and A\&B is our proposed auto-regressive two-step inference.}
	\label{figure3}
	\vspace*{-0.45cm}
\end{figure*}
\begin{figure}[t]
	\centerline{
 \subfloat[Transformer decoder]{{\includegraphics[width=1.35in,height=2in]{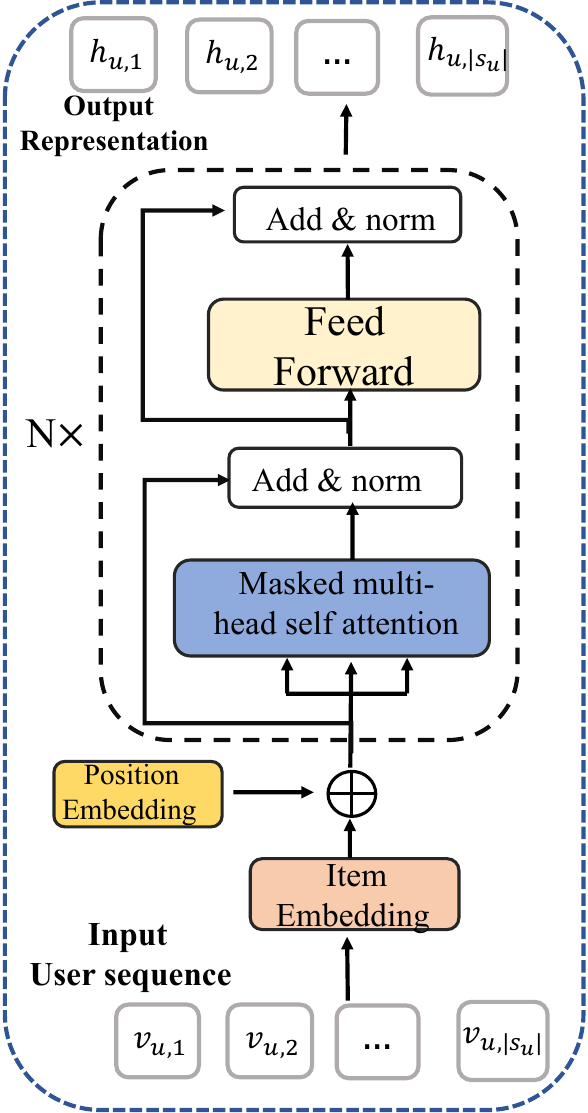}}}
 \ \
 \subfloat[Pre-training]{{\includegraphics[width=1.35in,height=2in]{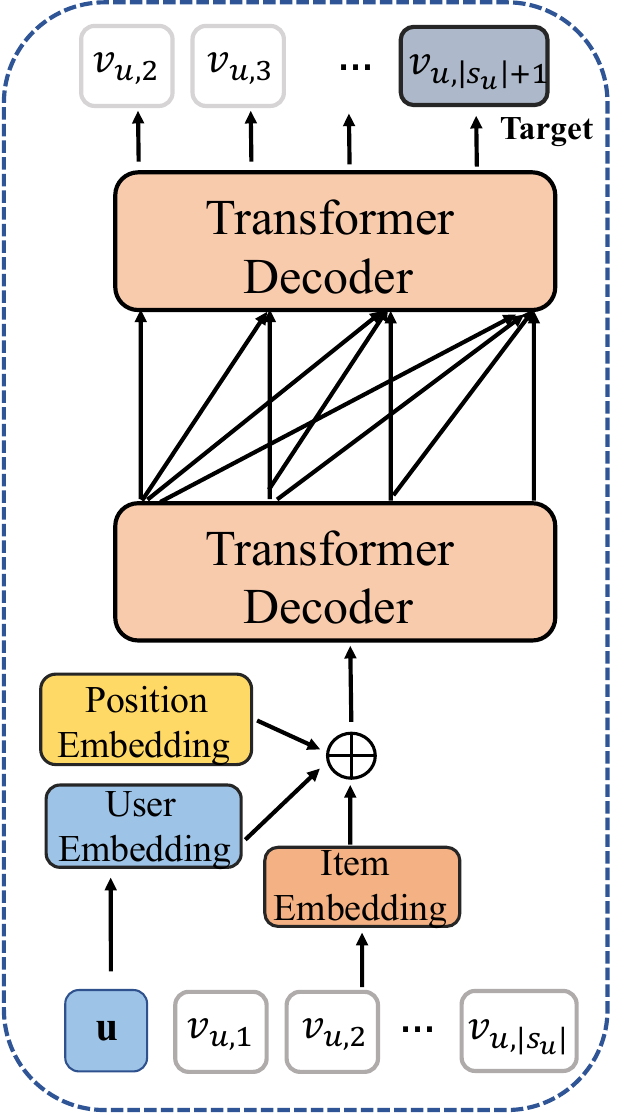}}}
 \ \
 \subfloat[Generative Prompts]{{\includegraphics[width=1.35in,height=2in]{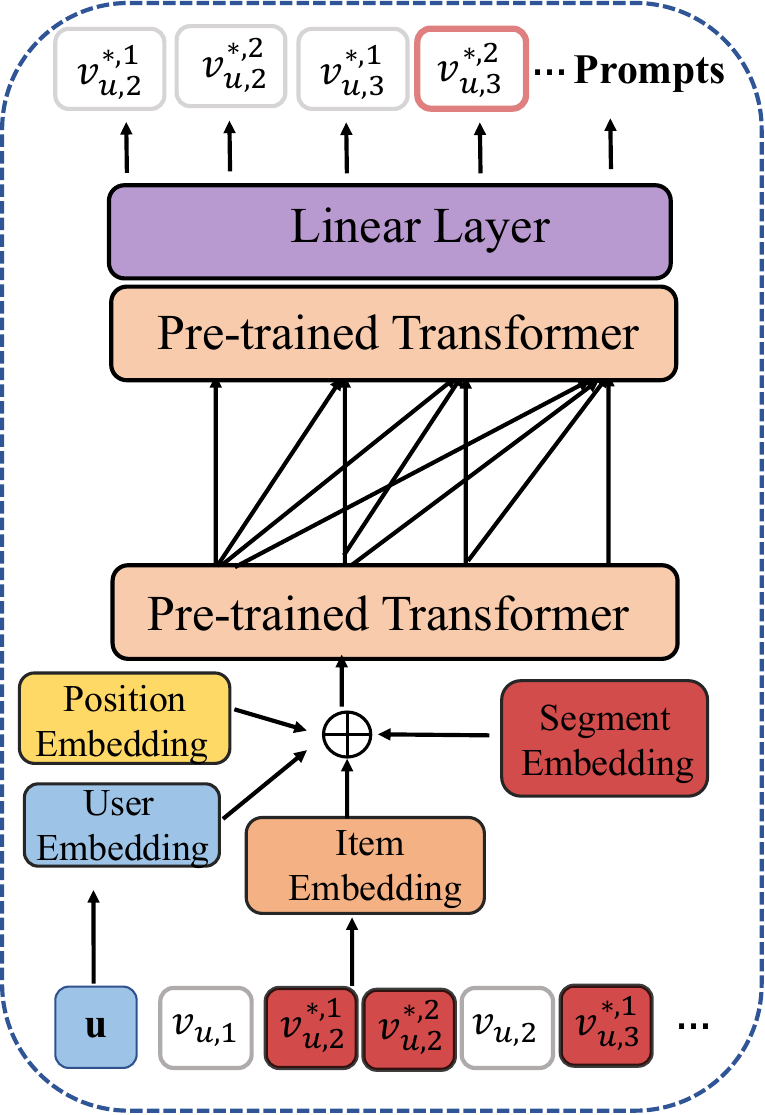}}}
 } 
 \vspace*{-0.2cm}
\caption{(a) Model: Generative Pre-training Transformer (GPT). (b) Pre-training framework based on GPT. (c) Fine-tuning framework with personalized prompts.}
	\label{figure2}
	\vspace*{-0.6cm}
\end{figure}
\section{Methodology}
\subsection{Overall Framework}
In this section, we present the details of our proposed RecGPT framework. Our training procedure consists of three stages, as illustrated in Fig. 2. The first stage is to pre-training a personalized auto-regressive generative model Transformer. This is followed by a Prompt-tuning stage, where we generate personalized prompts by conversations between the pre-trained model and users. Finally, we assess the capability of the auto-regressive recall during the inference stage.
\subsection{Pre-training auto-regressive generative model}
We first introduce our base model, Transformer decoder, to learn representations. 
The Transformer architecture \cite{vaswani2017attention} is a powerful method for encoding sequences. Various Transformer-based models \cite{kang2018self,sun2019bert4rec,xie2022contrastive} have been verified to be effective in sequential recommendation.
Following \cite{radford2018improving}, we also use the classical Transformer decoder as our pre-training model. As shown in Fig. 3(a), we stack multi-layer Transformer decoder blocks to decode the historical behavior sequence. For the input behavior sequence $s_{u}$, we define its $l$-th layer's behavior matrix as $h_{u}^{l}$=$\{h_{u,1}^{l},h_{u,2}^{l},\cdots,h_{u,|s_{u}|}^{l}\}$. The $l$-th layer encodes the previous layer's output $h_{l-1} $ into $h_{l} \in \mathcal{R}^{|s_{u}| \times d}$, where $l \in [1,n]$,$|s_{u}|$ is the length of the input sequence, and $d$ denotes the dimension item representations. 

Each layer applies a \textit{Masked Multi-head Self-attention} (MMS) module and a position-wise \textit{Feed-Forward Network} (FFN).  The latter is a two-layer FFN based on ReLU activation function. It performs linear transformations on the MMS's output $S_{l}  \in \mathcal{R}^{|s_{u}| \times d}$ and convents $S_{l}$ into $h_{l}$, $h_{l} =\text{ReLU}(S_{l}W_{l,1}+b_{l,1})W_{l,2}+b_{l,2}$,
where $W_{l,1} \in \mathcal{R}^{d \times d_{ff}}$, $b_{l,1} \in \mathcal{R}^{d_{ff}}$, $W_{l,2} \in \mathcal{R}^{d \times d_{ff}}$ and $b_{l,2} \in \mathcal{R}^{d_{ff}}$ are weight parameters.
The former MMS, which sub-layer aggregate $m$ attention heads, each of which computed identically with the scaled dot-product attention (\emph{e.g.,} the $h$-th head in the $l$-th layer $A_{l,h}$ $ \in  \mathcal{R}^{|s_{u}| \times  \frac{d}{m}}$). Formally, the computation of this sub-layer is defined as follows:
\begin{equation}
	\begin{split}
		&  S_{l} = [A_{l,1},\cdots,A_{l,m}] W^{S}_{l},\\
		&  A_{l,h} =\text{softmax}(\dfrac{Q_{l,h}K^{T}_{l,h}}{\sqrt{d}}+M)V_{l,h},\\
		&  S_{l} = h_{l-1}W^{Q}_{l,h}, \ h_{l-1}W^{K}_{l,h},\ h_{l-1}W^{V}_{l,h},\\
		& M=\left\{
		\begin{aligned}
			0 \ \  &\ \  \text{Allow to attend} \\
			-\infty &\ \  \text{Forbidden to attend} 
		\end{aligned}
		\right.
	\end{split}
\end{equation}
where $W^{S}_{l} \in \mathcal{R}^{d \times d}$, $W^{Q}_{l,h},W^{K}_{l,h},W^{V}_{l,h} \in \mathcal{R}^{d \times  \frac{d}{m}} $ are the projection matrices to be learned. $M \in \mathcal{R}^{|s_{u}| \times |s_{u}|}$ is the attention masking matrix and each element of $M$ controls whether an item in the sequence can be attended to another. The lower triangular part of M is set to $0$ and the remaining part $-\infty$, so as to allow each item to attend to past items, but prevent it from attending to future items. 

MMS models the item correlations in sequence. Position-wise FFN outputs a bag of embeddings, where the embedding at each position predicts the corresponding next item in the sequence. We present the next-item prediction of Pre-training in Fig. 3 (b). The $l$-th layer's behavior matrix  $h_{u}^{l}$ in our pre-training model is learned as follows:
\begin{equation}
	\begin{split}
		&h_{u}^{0} = uW_{u}+s_{u}W_{e} + W_{p},  \\
		&h_{u}^{l} = \text{Transformer\_block}(h_{u}^{l-1}), \\
		&h_{u,|s_{u}|} = f_{seq}(s_{u}|\theta) = h_{u,|s_{u}|}^{N},
	\end{split}
\end{equation}
Where $u$ is the user’s profile id, $W_{u}$ is the user embedding matrix, $s_{u}$ is the input behaviors sequence,  $W_{e}$ is the item embedding matrix, $W_{p}$ is the position embedding matrix, $h_{u,|s_{u}|}$ is the final user representation of $u$ learned in pre-training to predict the user's next item, $N$ is the number of Transformer decoder layers, and $\theta$ means all parameters. The main difference in our model compared to other models is that we design a user module to capture personalized information.

Given the above model, we then introduce the pre-training loss. The model $f_{seq}(\theta)$ is expected to estimate the next interacted item $v_{u, |s_{u}|+1}$. In practice, binary cross entropy is usually used to learn $f_{seq}$ based on the following loss:
\begin{equation}
	\begin{split}
		L_{SR}= -\!\!\sum_{u\in \mathcal U} \!\sum_{t\in [2,|s_{u}|]}\!\! \{\log(\sigma([h_{u,t}]^T \bm{e}_{u, t+1})) - \\ \!\!\!\!\!\!\!\sum_{s_{u, neg}\in O_{u,t}^-} \!\!\!\! \log(1-\sigma([h_{u, t}]^T \bm{e}_{u, neg})) \},
	\end{split}
\end{equation}
where $h_{u, t}$ is the representation of the sequence $\{v_{u,1},\cdots,v_{u,t}\}$ outputted by $f_{seq}$. $\delta$ is the non-linear activation function,
$O_{u,t}^-$ is the set of negative samples with respect to $v_{u,t+1}$.
$\bm{e}_{u,t+1}$ and $\bm{e}_{u,neg}$ are the embeddings of $v_{u,t+1}$ and the negative sample $s_{u,neg}$, respectively.
\vspace*{-0.3cm}
\subsection{Personalized Prompt-tuning}
\vspace*{-0.1cm}
After Pre-training, RecGPT utilizes a personalized Prompt-tuning based on conversations between GPT and users to perform next-item recommendation tasks, and the specific process of Prompt-tuning as shown in Fig. 2 (b).  

\textbf{Generative Personalized Prompts.} The key of our RecGPT is generating an effective prompt as feedback for user's request. However, it is challenging to find appropriate prompts in recommendation, since 1) it is difficult to build some real tokens as prompts. Unlike words in NLP, the tokens in recommendation do not have explicit meaningful semantics. 2) Furthermore, unlike NLP, recommendations should be personalized. Thus the prompts should also be customized for different users. In a sense, each user’s recommendation can be viewed as a task, while there are millions of users in a real-world system. It is impossible to design prompts manually for each user \cite{wu2022personalized}.  In RecGPT, we leverage user id and user behavior sequence item id to automatically and effectively construct personalized prompts for all users. The user id can represent specific attributes unique to each user. Our task of generating personalized prompts is closely connected to the auto-regressive generation approach used in unidirectional language models. As shown in Fig. 3(c),  $v^{i,*}_{u,j}$= $\{v^{1,*}_{u,2},v^{2,*}_{u,2},v^{1,*}_{u,3},\cdots \}$ is obtained personalized prompts for user behavior sequence $[v_{u,2}$, $v_{u,3},\cdots]$. New sequence $s^{*}_{u}$ with generative personalized prompts $v^{*,k}_{u,t+2}$ as depicted in Algorithm 1,  where the segment embedding matrix $W_{s}$ is designed to distinguish the generated prompts item id from the original behavior sequence item id, and the "argmax" is used to select item id of the highest probability over all items $\mathcal{V}$. 
\begin{algorithm}[t]
	\caption{Pseudocode to generate prompts sequence $s^{*}_{u}$} 
	\label{alg:algorithm}
	\textbf{Input}: User id $u$; Pre-trained "Transformer\_block"; User behavior sequence $s_{u} = \{v_{u,1},\cdots,v_{u,|s_{u}|} \}$; User embedding matrix : $W_{u}$; Item embedding matrix : $W_{e}$; Position embedding matrix : $W_{p}$; Segment embedding matrix : $W_{s}$; Linear output layer $W_{l}$; Whole items set $\mathcal{V}$; The number of personalized prompts generated: $K$.\\
	\textbf{Output}: New sequence $s^{*}_{u}$.
	\begin{algorithmic}[1] 
		\STATE Let $t=1$; $s^{*}_{u}=\{v_{u,1}\}$
		\WHILE{$t < |s_{u}|$}
		\STATE $k=1$
		\IF {$ k < K$}
		\STATE $h_{u}^{0}$= $uW_{u}$+$s^{*}_{u} W_{e}$ + $W_{p}$ + $W_{s}$;\\
		\STATE $h_{u}^{l}$ = $\text{Transformer\_block}(h_{u}^{l-1})$ \\
		\STATE $P_{u,t+1}$=$\text{softmax}(h_{u,|s_{u}|}^{N}W^{T}_{l})$\\
		\STATE $v^{*,k}_{u,t+2}$=$\text{argmax} \mathcal{V}(P_{u,t+1})$
		\STATE $s^{*}_{u} \leftarrow \text{Concatenation}(s^{*}_{u},v^{*,k}_{u,t+2})$
		\STATE $k = k+1$
		\ENDIF
		\STATE $s^{*}_{u} = [s^{*}_{u},v_{u,t+2}]$
		\ENDWHILE
		\STATE \textbf{return} $s^{*}_{u}$
	\end{algorithmic}
	\vspace{-0.1cm}
\end{algorithm}

\textbf{Prompt-tuning.} After generating the personalized prompts, we obtained a new prompt-enhanced sequence $s^{*}_{u}$: 
\begin{equation}
	\begin{split}
		s^{*}_{u} =  \{v_{u,1},\underline{v^{*,1}_{u,2},\cdots, v^{*,K}_{u,2}}, v_{u,2}, \underline{v^{*,1}_{u,3},\cdots, v^{*,K}_{u,3}},  v_{u,3},\cdots,\underline{v^{*,1}_{u,|s_{u}|\text{-}1},\cdots, v^{*,K}_{u,|s_{u}|\text{-}1}},v_{u,|s_{u}|} \} 
	\end{split}
	\nonumber
\end{equation}
where the generatived personalized prompts are indicated by \underline{underlining}. On the one hand, we need to train segment embedding matrix $W_{s}$ and an added linear output layer parameters $W_{l}$ except for the Pre-trained sequential model);  On the other hand, we also need to construct the final user representation for recommendation. Precisely, we first input the prompt-enhanced sequence  $s^{*}_{u}$ to the Pre-trained sequential model, and then we obtain the probability distribution over all items in the datasets. With the item's probability distribution, we then make the next-items prediction based on preceding items, which can be achieved by minimizing the following negative log-likelihood:
\begin{equation}
	\begin{split}
		L =  \frac{1}{|\mathcal{U}|}\sum_{u \in \mathcal{U} } &-\log p(v_{u,|s_{u}|+1} | s^{*}_{u,1}, s^{*}_{u,2+K}, s^{*}_{u,3+2K},\\&  \cdots, s^{*}_{u,|s_{u}|+(|s_{u}|-1)K}; \theta_{LP})
	\end{split}
\end{equation}
where $v_{u,|s_{u}|+1}$ is the next item to be predicted for user $u$, $K$ denotes the size of the generative personalized prompts window, and $\theta_{LP}$ represents all parameters as shown in Fig. 3 (c). 
\subsection{Inference-validating auto-regressive recall}
In the inference stage of sequential recommendation model, as shown in path \textbf{1} in Fig. 2 (c), existing methods mainly calculate the user's preference score for the item $i$ in the step (t+1) under the context from user history as:
\begin{equation}
	\begin{split}
		P(i_{t+1}=i|i_{1:t}) = e^{\mathrm{T}}_{i} \cdot F^{L}_{t},
	\end{split}
\end{equation}
where $e_{i}$ is the representation of item $i$ from the item embedding matrix $\text{M}_{I}$,  $F^{L}_{t}$ is the output of the L-layer Transformer decoder block at step $t$ and "$\cdot$" is the inner product. The method of path \textbf{1} is denoted as \textbf{RecGPT$_{1}$}. However, this inference mode can only predict the user's preferences for the next moment and cannot predict the user's preferences at multiple moments in the future.

In our generative personalized prompts model, we train the model using auto-regressive generative prompt items, enabling it to predict multiple future moments. Additionally, we investigate the effectiveness of auto-regressive inference in our model. Employing multiple rounds of auto-regressive recall can have a detrimental impact on the model's performance. On the one hand, sequential data is often sparse, making it challenging to train the model for auto-regressive recall without a substantial corpus. On the other hand, sequential behaviors usually exhibit the highest correlation between the two most recent items. However, as the number of auto-regressive recalls increases, the correlation between the recalled items diminishes progressively. Hence, we adopt a two-step auto-regressive recall approach in the inference stage. For example, when aiming to recall the top-k items, we implement a two-step strategy (\emph{i.e.,} $k=m+n$) represented by paths \textbf{A} and \textbf{B} in Fig. 2 (c). In path A, we initially retrieve the top-m items based on the output item representation  $h^{*}_{1}$ (where $i^{*}_{1}$ denotes the item ID) from the GPT-1 Pre-training model. Next, the  $i^{*}_{1}$ is concatenated with the original sequence before being fed into the Transformer decoder. The resulting representation, $h^{*}_{2}$, is then employed to recall the top-n items. This auto-regressive two-step recall strategy enables our model to recommend items that align with the user's multiple future preferences. We refer to this auto-regressive recall method as \textbf{RecGPT}. The main difference between RecGPT$_1$ and RecGPT is that the former relies on the traditional recall method, while the latter uses autoregressive generative recall method, as shown in Fig. 2(c).

\begin{table}[t]
	\begin{center}
		\renewcommand{\arraystretch}{1.1} 
		\caption{Statistics of the datasets after preprocessing.} 
 \vspace*{-0.01cm}
		\label{tab:freq1}
		\setlength{\tabcolsep}{0.95mm}{
			\begin{tabular}{c|c|c|c|c|c}
				\hline 
				Dataset& \#Users& \#Items& \#AveLen & Actions &Sparsity\\
				\hline
				Sports & 35,598& 18,357 &8.3 &296,337&99.95\%\\
				\hline
				Beauty& 22,363& 12,101 &8.9 &198,502&99.73\%\\
				\hline
				Toys  & 19,412  & 11,924 &8.6 &167,597 &99.93\%\\
				\hline
				Yelp& 30,431 & 20,033&10.3 &316,354&99.95\%\\
				\hline
		\end{tabular}}
	\end{center}
	\vspace{-0.5cm}
\end{table}
\begin{table*}[t]
	\begin{center}
		\renewcommand{\arraystretch}{0.85}
		\caption{Performance comparison of different methods. The best performing methods are \textbf{boldfaced} and the best baseline results are indicated by \underline{underline}.The improvements are significant under paired-t test (p $<$ 0.05).}
  \vspace{0.3cm}
		\label{tab:freq2}
		\setlength{\tabcolsep}{1.7mm} 
  \scalebox{0.9}{
		{\begin{tabular}{l|cccc|cccc}
				\hline  
				\ \ \multirow{3}{*}{\textbf{Model}}  & \multicolumn{4}{c|}{	\textbf{Sports and Outdoors}} & \multicolumn{4}{c}{	\textbf{Beauty}} \\
				& \multicolumn{2}{c}{\underline{\textbf{Metric@5}}}& \multicolumn{2}{c}{\underline{	\textbf{Metric@10}}}&  \multicolumn{2}{c}{\underline{	\textbf{Metric@5}}}& \multicolumn{2}{c}{\underline{	\textbf{Metric@10}}} \\
				& HR & NDCG &  HR & NDCG &  HR & NDCG&  HR & NDCG \\
				\hline
				PopRec & 0.0052& 0.0028 & 0.0081& 0.0037&  0.0064& 0.0030& 0.0098& 0.0040  \\
				BPR & 0.0030& 0.0019 & 0.0055& 0.0027& 0.0038& 0.0026& 0.0060& 0.0033    \\
				STAMP & 0.0079& 0.0051 & 0.0119& 0.0064& 0.0077& 0.0045& 0.0114& 0.0057    \\
				GRU4Rec & 0.0113& 0.0072   & 0.019& 0.0097 & 0.0162& 0.0097 &0.0300& 0.0141 \\
				NARM	 & 0.0132& 0.0085 & 0.0234& 0.0118 & 0.0230& 0.0142& 0.0401& 0.0197   \\
				$\text{S}^{3}$-$\text{Rec}_{\text{IPS}}$ & 0.0124& 0.0086 & 0.0205& 0.0111& 0.0202& 0.0119& 0.0336& 0.0163    \\
				BERT4Rec & 0.0200& 0.0130 & 0.0313& 0.0166 & 0.0382& 0.0210& 0.0592& 0.0319   \\
				SASRec & 0.0203& 0.0135 & 0.0324& 0.0174& 0.0398& 0.0261& 0.0614& 0.0331   \\
				Pre-train & 0.0203& 0.0132 & 0.0324& 0.0171 & 0.0418& 0.0272& 0.0610& 0.0334   \\
				Fine-tuning & \underline{0.0212}& \underline{0.0141} & \underline{0.0328}& \underline{0.0178}& \underline{0.0421}& \underline{0.0275}& \underline{0.0618}& \underline{0.0338}  \\
				\hline
				\textbf{RecGPT$_1$}	& 0.0213& 0.0141 & 0.0333& 0.0180 & 0.0426& 0.0283& 0.0651& 0.0356    \\
				\textbf{RecGPT}	& \textbf{0.0219}& \textbf{0.0143} & \textbf{0.0339}& \textbf{0.0181} & \textbf{0.0440}& \textbf{0.0289}& \textbf{0.0654}& \textbf{0.0357}  \\
				\textbf{Improved}	&  3.302\%&  1.418\%&  3.354\%&  1.685\%&  4.513\%&  5.091\%&  5.825\%&  5.621\% \\
				\hline
				\hline
				& \multicolumn{4}{c|}{	\textbf{Toys and Games}} & \multicolumn{4}{c}{	\textbf{Yelp}}  \\
				\hline
				PopRec & 0.0039& 0.0021 & 0.0070& 0.0031 & 0.0045& 0.0023& 0.0091& 0.0038  \\
				BPR & 0.0030& 0.0019& 0.0047& 0.0024 & 0.0023& 0.0015& 0.0039& 0.0021   \\
				STAMP & 0.0033& 0.0019 & 0.0061& 0.0028 & 0.0041& 0.0025& 0.0071& 0.0035  \\
				GRU4Rec & 0.0157& 0.0097 & 0.0274& 0.0134& 0.0122& 0.0076& 0.0219& 0.0107  \\
				NARM	& 0.0257& 0.0174 & 0.0405& 0.0221 & 0.0152& 0.0095& 0.0256& 0.0129  \\
				$\text{S}^{3}$-$\text{Rec}_{\text{IPS}}$ & 0.0227& 0.0146 & 0.0397& 0.0200 & 0.0159& 0.0094& 0.0269& 0.0129   \\
				BERT4Rec & 0.0455& 0.0307 & 0.0683& 0.0380&   0.0156& 0.0096& 0.0262& 0.0130 \\
				SASRec & 0.0479& 0.0334 & 0.0698 & \underline{0.0405} & 0.0161& 0.0100& \underline{0.0290}& \underline{0.0142} \\
				Pre-train & 0.0487&  \underline{0.0342} & 0.0630& 0.0388 & 0.0166& 0.0102& 0.0278& 0.0138 \\
				Fine-tuning & \underline{0.0492}& 0.0335 & \underline{0.0701}& 0.0402 & \underline{0.0168}& \underline{0.0103}& 0.0287& 0.0141   \\
				\hline
				\textbf{RecGPT$_1$}	&0.0515&	0.0355&	0.0711&	0.0418	& 0.0172& 0.0105 &0.0293& 0.0144  \\
				\textbf{RecGPT}	&\textbf{0.0529}&	\textbf{0.0359}&	\textbf{0.0721}&	\textbf{0.0419}&  \textbf{0.0177}& \textbf{0.0107}& \textbf{0.0297}&\textbf{0.0146}	\\
				\textbf{Improved}	&  7.520\%&  4.971\%&  2.853\%&  3.457\%&  5.357\%&  3.883\%&  3.484\%&  3.546\% \\
				\hline
		\end{tabular}}}
	\end{center}
	\vspace{-0.6cm}
\end{table*}

\section{Experiments}
 \vspace*{-0.1cm}
To answer the following questions, we conduct experiments to evaluate the effectiveness of the proposed model with comparison to state-of-the-art methods.  
\begin{itemize}
	\item \textbf{RQ1:} How does RecGPT perform compared with previous approaches? 
	\item \textbf{RQ2:} How do hyper-parameters influence the results of RecGPT? 
	\item \textbf{RQ3:} How do different components affect the results of RecGPT?
\end{itemize}
 \vspace*{-0.5cm}
\subsection{Experimental Setup}
 \vspace*{-0.1cm}
\noindent
\textbf{Datasets:} In order to verify the effectiveness of our model. We conduct experiments on four public datasets collected from two real-world platforms. After prepossessing, the statistics of these datasets are summarized in Table \uppercase\expandafter{\romannumeral1}.

\textbf{Sports and Outdoors},  \textbf{Beauty} and \textbf{Toys and Games}: these three \footnote{http://jmcauley.ucsd.edu/data/amazon/} datasets are obtained from Amazon review datasets in \cite{mcauley2015image}. 
\textbf{Yelp}\footnote{https://www.yelp.com/dataset}: this is a popular dataset for business recommendation. Following \cite{xie2020contrastive}, we only use the transaction records after January 1st, 2019.  For all datasets, we follow common practice in \cite{zhou2020s3} to preprocess the datasets. Specificity, the numeric ratings or the presence of a review are treated as positive instances while others as negative instances. We keep the ‘5-core’ datasets, i.e., every user bought in total at least 5 items and each item were bought by at least 5 users. 

\vspace{0.2cm}
\noindent
\textbf{Baselines:} To fully evaluate the performance of our methods, we compare our method with following classic recommendation methods: 

$\bullet$ \textbf{{Non-sequential models}} include PopRec and BPR \cite{rendle2012bpr} . 

$\bullet$ \textbf{{Sequential models}} include STAMP \cite{liu2018stamp}, GRU4Rec \cite{hidasi2015session}, NARM \cite{li2017neural}, $\text{S}^{3}$-$\text{Rec}_{\text{IPS}}$ \cite{zhou2020s3}, BERT4Rec \cite{sun2019bert4rec} and SASRec \cite{kang2018self}. 

$\bullet$ \textbf{{Foundation models}} include Pre-train and Fine-tuning. \textbf{Pre-train:} We directly conduct the pre-trained model on the test set as a baseline for comparison. \textbf{Fine-tuning:} It is a strong tuning method to utilize pre-trained models for SR tasks. It tunes all pre-trained models parameters in tuning with the same user id and historical behaviors.

\vspace{0.2cm}
\noindent
\textbf{Implementation Details:} for each baseline model, we implement them by PyTorch in RecBole \cite{recbole2.0}. Note that RecBole uses data augmentation by default. To ensure fairness, following \cite{kang2018self}, turned off this feature and compared the all baselines in the same data processing instead. For example, if the original user sequence is ABCDE, then the generated samples are AB $\rightarrow$ C, ABC $\rightarrow$ D and ABCD $\rightarrow$ E. For the Pre-training stage, we only use samples AB $\rightarrow$ C. For the Prompt-tuning stage and other baseline models, we adopt the leave-one-out strategy to evaluate the performance, \emph{i.e.}, AB $\rightarrow$ C and ABC $\rightarrow$ D are used for model training and validation, while the last is left for testing. Our method is implemented in PyTorch. For the Pre-training model Transformer encoder architecture, we set self-attention blocks as 1 and attention heads as 2, the dimension of the embedding as 64, and the maximum sequence length as 50. The model is optimized by an Adam optimizer with a learning rate 0.001 and batch size as 256. For hyper-parameters of our methods, we tune the size of generative personalized prompts window $K$ in ranges of [0,1,$\cdots$,6]. The number of two-step auto-regressive generative recall items, $m$ and $n$, ranges from 1 to 20. In the offline experiment, during the Prompt-tuning stage, we use the last 50 items of the newly generated sequence to keep the same sequence length.

\noindent
\textbf{Metrics:} 
For overall evaluation, we employ top-k Hit Ratio (HR@k) and top-k Normalized Discounted Cumulative Gain (NDCG@k) \cite{jarvelin2002cumulated} with k $\in$ $\{$5, 10$\}$. We evaluate the ranking results over the whole item set for a fair comparison \cite{qiu2022contrastive}.

\subsection{Performance Comparison (RQ1)}
\vspace*{-0.4cm}
The results of all the methods are reported in Table 2. For the results, we have the following observations.

$\bullet$  The PopRec and BPR non-sequential model performs inferiorly when compared to all other sequential models, suggesting that mining user behavior sequences is an effective method for recommending the next item to the user.

$\bullet$ In sequential models, SASRec utilizes a Transformer based encoder and consistently performs better than non-Transformer methods such as GRU4Rec, Caser, NARM, and STAMP, highlighting the effectiveness of the multi-head self-attention mechanism for sequential recommendation. BERT4Rec and S$^{3}$-Rec$_{\text{IPS}}$
are trained with an additional supervisory signal. BERT4Rec inherits an attention-based encoder while introducing additional objectives. S$^{3}$-Rec$_{\text{IPS}}$ is proposed for the fusion of additional contextual information. However, in most cases, we observed that both methods perform worse than SASRec. This performance difference can be attributed to BERT4Rec's strong emphasis on context information in user behavior sequences and S$^{3}$-Rec$_{\text{IPS}}$'s two-stage training strategy, which hinders information sharing between the next-item prediction and SSL tasks, leading to inferior results.

$\bullet$ Among the two methods proposed by us, out of statistical tests, our personalized prompt-based model RecGPT$_1$ (as denoted in section 4.4) and RecGPT outperformed the best counterpart baselines about 88\% and 96\% cases, respectively. It indicates that our personalized prompts can better extract useful historical behavior information related to the current user from the pre-training models, which is demonstrating its effectiveness for sequential recommendation. Furthermore, the results suggest that generative personalized prompt methods are more effective than traditional sequential recommendation methods. Interestingly, the RecGPT model consistently outperforms the RecGPT$_1$ model in all cases. This finding suggests that utilizing an auto-regressive recall method during the inference stage effectively captures a user's future preferences across multiple time points. In general, we present a novel training paradigm based on ChatGPT, and experimental results demonstrate its effectiveness.

\begin{figure}[t]
	\centerline{
		{\includegraphics[width=1.7in,height=1.7in]{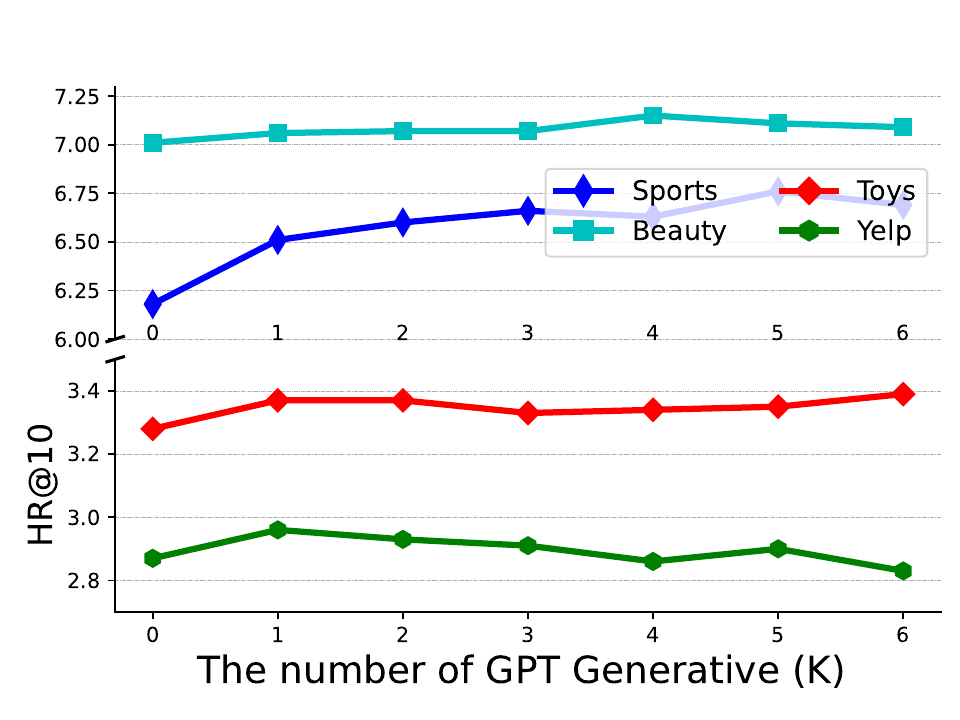}}
		\ \ \ \ \  \ \ 
		{\includegraphics[width=1.7in,height=1.7in]{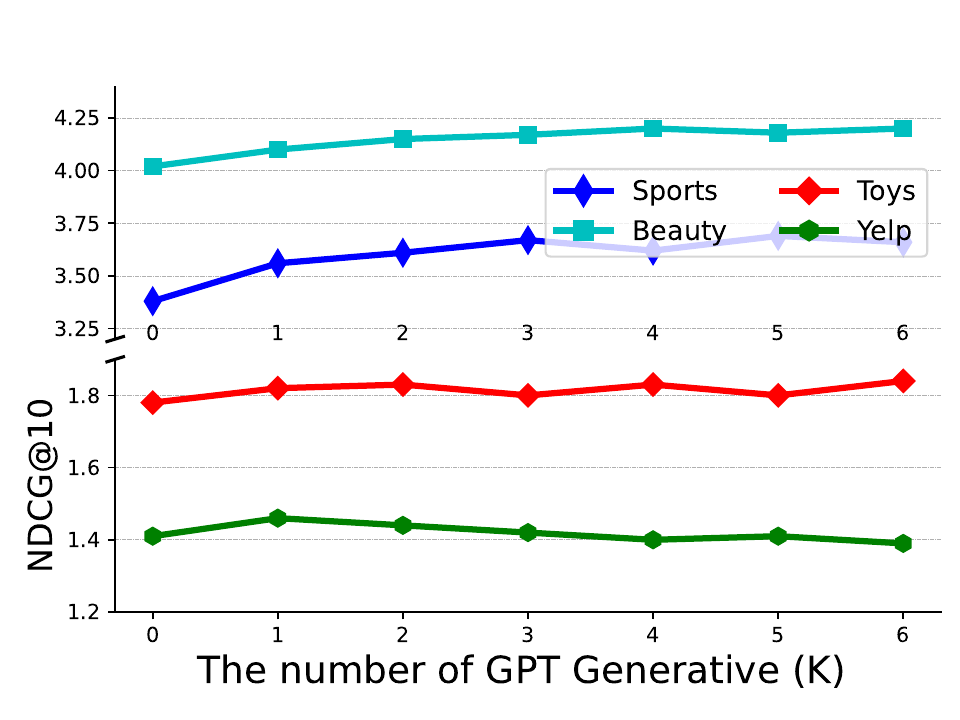}}
	} 
	\caption{Sensitive of the size of window $K$ in term of HR@10 and NDCG@10.}
	\label{figure4}
	\vspace*{-0.4cm}
\end{figure}
\vspace{-0.3cm}
\subsection{Parameter Analysis (RQ2)}
This section examines the impact of key hyper-parameters on the performance of our framework RecGPT. The results are reported for all four datasets according to the HR@10 and NDCG@10 metrics. All the numbers are percentage values with “\%” omitted. The hyper-parameters include  $K$ and $m\_n$. 

\textbf{Influence of the hyper-parameters $K$:} The parameter $K$, which controls the size of the generative personalized prompts window, and we tune it within the range of [0, 6]. As shown in Fig. 4, increasing the number of $K$ enhances the model's performance, due to the effective similarity preference matching of the personalized prompts, and the best performance is obtained when the window size is in the range [1, 3]. The reason may be that too many steps will generate some irrelevant prompts, which will weaken the correlation of the original sequence. 

\textbf{Influence of the hyper-parameter $m\_n$:} The parameter $m\_n$ controls the ratio of two-step recall. Due to page limitations, we only report results for ratios ranging from (10\_0) to (5\_5) at top@10. Analysis of Fig. 5 and 6 indicates optimal performance when $m_n$ is set to (9,1) and (8,2). This is attributed to the sparsity of sequence data, where auto-regressive recall requires a substantial amount of data. Additionally, the sequence data exhibits the strongest connection between the two behavior items, making the first-step generation of representation recall more crucial than the second step.

\begin{figure}[t]
	\centerline{
 \scalebox{0.65}{{\includegraphics[width=7.5in,height=1.5in]{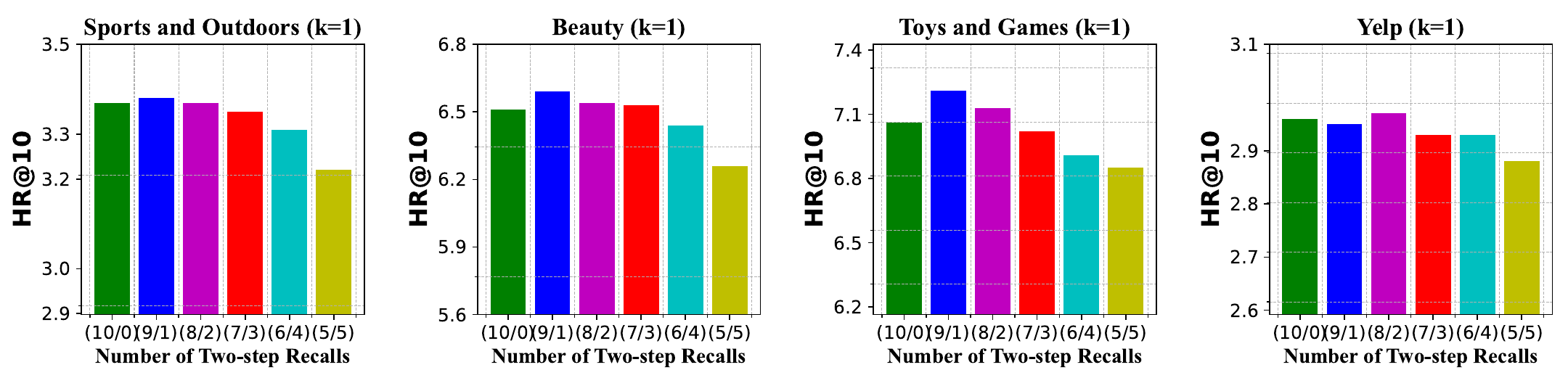}}}}
 \centerline{
	\scalebox{0.65}{\centerline{{\includegraphics[width=7.5in,height=1.5in]{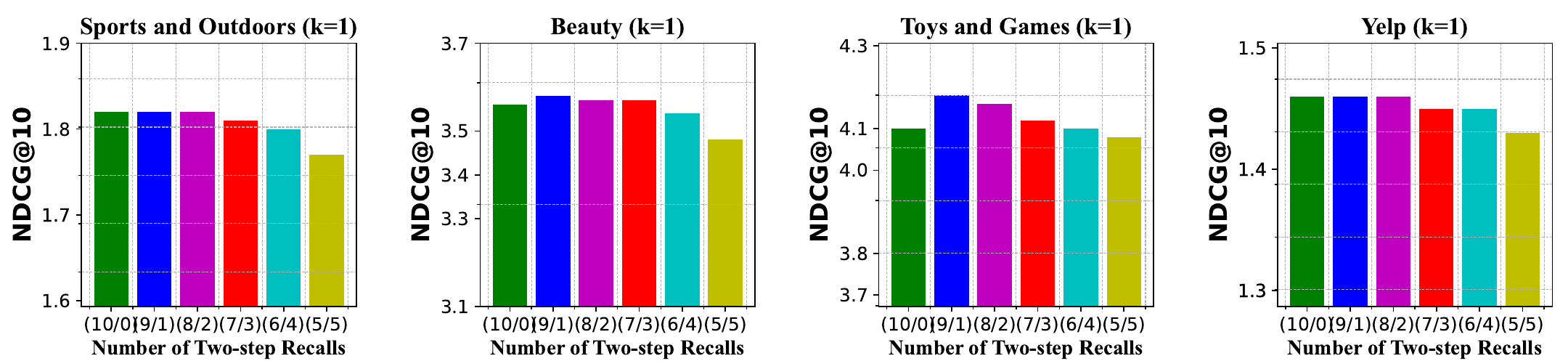}}}}} 
	\caption{Sensitive of parameter $m\_n$ in term of HR@10 and NDCG@10.}
	\label{figure5}
	\vspace*{-0.4cm}
\end{figure}
\vspace*{-0.3cm}
\subsection{Ablation Study (RQ3)}
\vspace*{-0.1cm}
We conduct an ablation study to investigate the contribution of each component, where the results are shown in Table 3:

\textbf{Variant\_1:}  We evaluate the two-step recall using the Pre-training model directly, without applying Prompt-tuning. In all cases, its performance is significantly inferior to that of RecGPT, underscoring the importance of Prompt-tuning component. \textbf{Variant\_2:} We replace two-step recall with the traditional inner product for evaluation (\emph{i.e.,} RecGPT$_1$ as denoted in section 4.4). The result indicates the effectiveness and superiority of two-step auto-regressive recall.  \textbf{Variant\_3:} We remove both Prompt-tuning and two-step recall. 
Comparisons between RecGPT and Var\_3 shows that Var\_3 performs worse than RecGPT in all cases, indicating the importance of the above components. Furthermore, the performance of Var\_3 is better than Var\_1, which further shows that auto-regressive recall directly in the infer stage without Prompt-tuning will damage the model income. 
\vspace{-0.1cm}
\subsection{Online A/B Testing (RQ1)}
We further tested of our approach through a live A/B experiments conducted on the \textit{Kuaishou} video APP recommendation platform. 
\begin{table}[t]
	\renewcommand{\arraystretch}{0.9} 
	\caption{Performance of ablation on different components.} 
    \vspace{0.1cm}
\label{PLL_realword_table}
\centering
\setlength{\tabcolsep}{0.6mm}
{ \begin{tabular}{clcccc}
		\hline 
		\textbf{Datasets}& \textbf{Metric}& \textbf{RecGPT}& \textbf{Variant\_1} & \textbf{Variant\_2}  & \textbf{Variant\_3} \\
		\hline
		\multirow{2}{*}{Sports} & HR@10 &   \textbf{0.0339}& 0.0226& 0.0333& 0.0324\\
		&NDCG@10 &   \textbf{0.0181}&0.0135 & 0.0180& 0.0171\\
		\hline
		\multirow{2}{*}{Beauty} & HR@10 & \textbf{0.0654}& 0.0560& 0.0651& 0.0610\\
		&NDCG@10 &   \textbf{0.0357}& 0.0315& 0.0356& 0.0334\\
		\hline
		\multirow{2}{*}{Toys} & HR@10 &   \textbf{0.0721}& 0.0603 & 0.0711& 0.0630\\
		&NDCG@10 &   \textbf{0.0419}& 0.0364& 0.0415& 0.0388\\
		\hline
		\multirow{2}{*}{Yelp} & HR@10 &   \textbf{0.0297}& 0.0261& 0.0293& 0.0278\\
		&NDCG@10 &   \textbf{0.0146}& 0.0133& 0.0144& 0.0138\\
		\hline
\end{tabular}}
\vspace{-0.4cm}
\end{table}

\textbf{Experimental setup:} 
We use our method to replace the baseline retrieval of the current \textit{Kuaishou} recommender system APP. Specifically, when a user arrives, the recommender system first generates a set of candidate videos that the user might be interested in based on his/her characteristics (\emph{a.k.a} the retrieval phase). We use auto-regressive method to generate six user\_embeds for the user and use inner product as the measurement standard for parallel ANN retrieval. We conduct A/B experiments to evaluate their live performances. 

\textbf{Compared methods:} 
Our online baseline method is "User-Aware Multi-Interest Learning with ComiRec \cite{cen2020controllable}", which is the current top-ranked main recall method of \textit{Kuaishou} video APP.

\textbf{Metrics:} 
The model's performance is evaluated based on the consumption metrics of users, we consider a broader range of other indicators including \textbf{comment} (leaving comments on videos), \textbf{forward} (forward videos with friends), \textbf{play} (video play count), \textbf{follow} (following video creators) and \textbf{watch time}. 

\textbf{Results:} 
From 7/1/2023 to 7/5/2023, online A/B testing was conducted in the recalling system of  \textit{Kuaishou} APP contributes +0.772\% \textbf{Comment}, +0.336\% \textbf{Forward}, +0.143\% \textbf{Play}, +0.027\% \textbf{Follow} and +0.017\% \textbf{Watch time}  gain compared to the previous ComiRec model online.

\vspace{-0.2cm}
\section{Conclusion}
\vspace{-0.1cm}
In this work, we explore a novel ChatGPT-based training paradigm for sequential recommendation. we design a new chat framework in item index level for the recommendation task, which includes three parts: model, training and inference. 
The results from experiments on four offline public datasets and an online A/B testing experiment all confirm the effectiveness of generative personalized prompts in the sequential recommendation task. In future research, we plan to introduce reinforcement strategies in public datasets to manually annotate real feedback annotations for personalized prompts on our proposed model.

\vspace{0.5cm}
\noindent
\textbf{Acknowledgment}. This work is supported in part by National Key R\&D Program of China (2023YFF0905402),National Natural Science Foundation of China (No. 62102420), Beijing Outstanding Young Scientist Program NO. BJJWZYJH\\012019100020098, Intelligent Social Governance Platform, Major Innovation \& Planning Interdisciplinary Platform for the ”Double-First Class” Initiative, Renmin University of China, Public Computing Cloud, Renmin University of China, fund for building world-class universities (disciplines) of Renmin University of China, Intelligent Social Governance Platform.

 \bibliographystyle{splncs04}
 \bibliography{splncs}
%
%
%
%
%
\end{document}